\documentstyle[12pt,epsfig]{article}
\begin{document}
\noindent
\begin{center}
{\Large {\bf Constraining $f(R)$ Theories with Temporal Variation
of Fine Structure Constant\\}} \vspace{2cm}
 ${\bf Yousef~Bisabr}$\footnote{e-mail:~y-bisabr@srttu.edu.}\\
\vspace{.5cm} {\small{Department of Physics, Shahid Rajaee Teacher
Training University,
Lavizan, Tehran 16788, Iran.}}\\
\end{center}
\vspace{1cm}
\begin{abstract}
It is well-known that $f(R)$ theories in Einstein frame is
conformally equivalent to quintessence models in which the scalar
field minimally couples with gravity.  If there exists a matter
system in Jordan frame, then it interacts with the scalar field in
Einstein frame due to the conformal transformations. This
interaction, in general, may lead to changes of fundamental
constants. Here we will consider possible time variation of fine
structure constant in a general $f(R)$ theory. We will use
observational bounds on these variations and argue that it
provides a criterion for constraining $f(R)$ models.

\end{abstract}
\section{Introduction}
Recent observations on expansion history of the universe indicate
that the universe is experiencing a phase of accelerated expansion
\cite{sup}.  This can be interpreted as evidence either for
existence of some exotic matter components or for modification of
the gravitational theory.  In the first route of interpretation
one can take a perfect fluid with a sufficiently negative
pressure, dubbed dark energy \cite{mel}, to produce the observed
acceleration. In the second route, however, one attributes the
accelerating expansion to a modification of general relativity. A
particular class of models that has recently drawn a significant
amount of attention is the so-called $f(R)$ gravity models
\cite{car}\cite{r}. These models propose a modification of
Einstein-Hilbert action so that the scalar curvature is replaced
by some arbitrary function $f(R)$.\\ It is well known that
conformal transformations can be used to recast $f(R)$ theories of
gravity into the form of Einstein gravity together with a
minimally coupled scalar field. The original variable is called
Jordan conformal frame while the transformed set, whose dynamics
is described by Einstein equations, is called Einstein conformal
frame.  If one formulates a $f(R)$ theory with some matter systems
in the Jordan frame, then the conformal transformation induces a
coupling of the scalar field with the matter sector in the
Einstein frame. One of the important implications of such a
coupling is the possible changes of constants of nature. If the
effective mass of the scalar field is sufficiently small the
interaction of the field
 with ordinary matter results in the time variation of constants of nature over
 cosmological time scales \cite{carrol}.  Among all possibilities, we restrict
ourselves to time variation of fine structure constant as a
consequence of interaction of the scalar field with
electrodynamics. We intend here to use the current
 bounds on a running fine structure constant to constrain a
 general $f(R)$ theory in the Einstein frame.

~~~~~~~~~~~~~~~~~~~~~~~~~~~~~~~~~~~~~~~~~~~~~~~~~~~~~~~~~~~~~~~~~~~~
\section{Quintessence and time variation of fine structure constant}
Many years after Dirac's proposal \cite{dirac} that fundamental
constants may vary with time and/or space, there is now some
theoretical frameworks which predict such changes.  For instance,
one of the most interesting low energy features of string theory
is the possible presence of a massless scalar field, dilaton or
moduli fields, whose vacuum expectation values define the size of
the effective coupling constants.  Independent of this framework,
Bekenstein \cite{beck} has also formulated a theory for varying
fine structure constant which has been recently generalized and
applied to a cosmological setting \cite{ba}\cite{oli}. Moreover,
there is a large class of dark energy and quintessence models
which invoke scalar fields with wavelengths comparable to the size
of the universe. This scalar field may then interact with matter
and radiation so that changes of its background value induce
variation of coupling constants \cite{carrol}. \\In recent years,
there is an increasing interest to models concerning time
variation of fine structure constant due to a non-vanishing
coupling of quintessence to the electromagnetic field.  In these
models, one can combine cosmological data and terrestrial
observations to place empirical constraints on the size of the
coupling or even explore the conditions that the scalar field
exhibit a tracking behavior \cite{cope}\cite{l}\cite{e}. The
general form of the action in these models is the following
\footnote{We work in the unit system in which $\hbar=c=8\pi G=1$
and our sign convention is (-+++).}
\begin{equation}
S=\frac{1}{2}\int d^4 x \sqrt{-g} R +\int d^4 x \sqrt{-g}
(L_{\phi}+L_{m}+L_{\phi F}) \label{1}\end{equation} where $R$ is
the curvature scalar and $L_{m}$ is Lagrangian density of cosmic
matter which we take it as a perfect fluid with energy density
$\rho_{m}$ and equation of state parameter $\omega$,  and
\begin{equation}
L_{\phi}=\frac{1}{2}\partial^{\mu}\phi \partial_{\mu} \phi-V(\phi)
\end{equation}
\begin{equation}
L_{\phi F}=-\frac{1}{4}A(\phi)F_{\mu\nu}F^{\mu\nu}
\end{equation}
$A(\phi)$ is a dimensionless function\footnote{We may call it a
gauge function since evolution of fine structure constant strongly
depends on the choice of this function.} of $\phi$ and
$F_{\mu\nu}$ is the components of electromagnetic field tensor.
For spatially flat Friedmann-Robertson-Walker (FRW) spacetime, the
governing field equations for the scale factor $a(t)$ and the
scalar field are
\begin{equation}
H^2=\frac{1}{3}(\rho_{m}+\frac{1}{2}\dot{\phi}^2+V(\phi))
\end{equation}
\begin{equation}
\dot{H}=-\frac{1}{2}(\omega \rho_{m}+\dot{\phi}^2)
\end{equation}
\begin{equation}
\ddot{\phi}+3H\dot{\phi}+\frac{dV(\phi)}{d\phi}=0
\end{equation}
where $H\equiv \frac{\dot{a}}{a}$. These field equations depend on
the choice of the functions $V(\phi)$ and $A(\phi)$.  These
functions characterize the evolution of the scalar field and the
effective fine structure constant which is now given by
\begin{equation}
\alpha(\phi)=\frac{\alpha_{0}}{A(\phi)} \label{al}\end{equation}
with $\alpha_{0}$ being its present value and $A(\phi_{0})=1$. The
choice of the gauge function $A(\phi)$ is arbitrary although the
choice of linear \cite{l} and exponential \cite{ba}\cite{e}
functions are usual in the literature. It should give an
appropriate temporal variations of fine structure constant which
is consistent with observational constraints. We bring in the
following two of these constraints which we will use in the next
section\footnote{For a more
complete list of these constraints see, e.g., \cite{e} and references therein.} :\\
1) Estimates of the age of iron meteorites ($z=0.45$) combined
with a measurement of the ratio Re/Os resulting from the decay
rate $R^{187}\rightarrow Os^{187}$ gives  \cite{14}
\begin{equation}
\mid \frac{\triangle \alpha}{\alpha} \mid < 10^{-6}
\label{a2}\end{equation} 2) We have also \cite{0}
\begin{equation}
\mid \frac{\dot{\alpha}}{\alpha} \mid < 4.2 \times 10^{-15}~
yr^{-1} \label{a3}\end{equation} coming from comparing atomic
clocks at present time ($z=0$). The overdot denotes
differentiation with respect to cosmic time.
~~~~~~~~~~~~~~~~~~~~~~~~~~~~~~~~~~~~~~~~~~~~~~~~~~~~~~~~~~~~~~~~~~~~~~~~~~~~~~~~~~~~~~~~~~~~~~~~~
\section{$f(R)$ Gravity}
The action of $f(R)$ gravity in Jordan frame in the presence of
matter is given by
\begin{equation}
S=\frac{1}{2}\int d^4 x \sqrt{-g} f(R) + \int d^4 x
\sqrt{-g}~L_{m}(g_{\mu\nu}, \psi)\label{a1}\end{equation} where
$f(R)$ is an arbitrary function of the scalar curvature. The
Lagrangian density $L_{m}$ corresponds to matter fields which are
collectively denoted by $\psi$.  We apply the following conformal
transformation
\begin{equation}
\tilde{g}_{\mu\nu}=\Omega~g_{\mu\nu}~,~~~~~~~~~~~~~~~~~~~~~~~~\Omega=f^{'}(R)
\label{a5}\end{equation}  This together with a redefinition of the
conformal factor in terms of a scalar field
\begin{equation}
\phi=\frac{1}{2\beta} \ln \Omega \label{o}\end{equation} with
$\beta=\sqrt{\frac{1}{6}}$, yields \cite{soko}
\begin{equation}
S=\frac{1}{2}\int d^4 x ~\sqrt{-\tilde{g}}~ \{
\tilde{R}-\tilde{g}^{\mu\nu}\nabla_{\mu}\phi\nabla_{\nu}\phi-V(\phi)\}
+\int d^4 x \sqrt{-\tilde{g}}~e^{-4\beta
\phi}~L_{m}(\tilde{g}_{\mu\nu}, \psi)
 \label{a6}\end{equation}
 In
the action (\ref{a6}), $\phi$ is minimally coupled to
$\tilde{g}_{\mu\nu}$ and appear as a massive self-interacting
scalar field with a potential
\begin{equation}
V(\phi)=\frac{1}{2}e^{-2\beta
\phi}r[\Omega(\phi)]-\frac{1}{2}e^{-4\beta \phi}
f(r[\Omega(\phi)]) \label{a7}\end{equation} where the function
$r(\Omega)$ is the solution of the equation
$f^{'}[r(\Omega)]-\Omega=0$ \cite{soko}. Thus the variables
$(\tilde{g}_{\mu\nu}, \phi)$ provide the Einstein frame variables
for $f(R)$ theories.  This Einstein frame representation is
equivalent to a quintessence model with a specific potential
function which, due to (\ref{a7}), is closely
related to the function $f(R)$.\\
It is important to note that in the action (\ref{a6}) the scalar
field interacts with matter sector via the function $e^{-4\beta
\phi}$. If we take $L_{m}$ to be the Lagrangian density of
electromagnetic field, then the action (\ref{a6}) will become
similar to (\ref{1}) with a gauge function $A(\phi)=e^{-4\beta
\phi}$. Thus one can generally state that any $f(R)$ theory in the
Einstein frame is equivalent with a quintessence model in which
the scalar field interacts with matter sector via an exponential
gauge function. However, it should be pointed out that in the
model (\ref{1}) evolution of the scalar field depends on the
choice of two arbitrary functions $A(\phi)$ and $V(\phi)$ while in
the model (\ref{a6}) it depends only on $V(\phi)$ which in turn is
characterized by the function $f(R)$. It implies that depending on
cosmological evolution of $\phi$, or equivalently on functional
form of $f(R)$, the fine structure constant could have many
possible histories during evolution of the universe. It is
therefore worth studying what possibilities are allowed by the
function $f(R)$ within the available observational constraints.
 \\
To proceed furthur, we note that (\ref{al}) is equivalent to
$\alpha=\alpha_{0}~e^{4\beta \phi}$. We would like to use
(\ref{a5}) and (\ref{o}) to write this relation in terms of scalar
curvature rather than the scalar field, namely that
$\alpha=\alpha_{0}~f'^{2}(R)$ where the prime denotes
differentiation with respect to the argument. Thus the rate of
change of $\alpha$ is given by
\begin{equation}
\frac{\dot{\alpha}}{\alpha}=2\dot{R}\frac{f''(R)}{f'(R)}
\label{a9}\end{equation} In a spatially flat FRW spacetime
$R=6(\dot{H}+2H^2)$ and therefore $\dot{R}=6(\ddot{H}+4\dot{H}H)$.
In terms of deceleration parameter $q\equiv
-a\ddot{a}/\dot{a}^2=-(1+\dot{H}/H)$ and jerk $j\equiv a^2
\dot{\ddot{a}}/\dot{a}^3$ we will have
\begin{equation}
\frac{\dot{\alpha}}{\alpha}=12(j-q-2)H^3 \frac{f''(R)}{f'(R)}
\label{a10}\end{equation} It is also possible to obtain the
relative change of $\alpha$,
\begin{equation}
\frac{\triangle \alpha}{\alpha} \equiv
\frac{\alpha(z)-\alpha_{0}}{\alpha_{0}} \label{a11}\end{equation}
which is equivalent to
\begin{equation}
\frac{\triangle \alpha}{\alpha}=f'^2(R)\mid_{z}-1
\label{a12}\end{equation} In this relation  $\alpha(z)$ and
$f'^2(R)\mid_{z}$ indicate fine structure constant and $f'^2(R)$
at redshift $z$.  We can now use (\ref{a10}) and (\ref{a12}) to
study the impact of the constraints (\ref{a2}) and (\ref{a3}) on a
particular $f(R)$ model. \\Let us first consider the model
\cite{amen}
\begin{equation}
f(R)=R+\lambda R_{c}(\frac{R}{R_{c}})^n \label{R}\end{equation}
where $\lambda$ and $n$ are positive parameters and $R_{c}$ is of
the order of the presently observed effective cosmological
constant. Here and in the following we will take
$R_{c}=\varepsilon H_{0}^2$ with $\varepsilon$ being a constant of
order of unity and $H_{0}$ is the Hubble constant. We substitute
this $f(R)$ function into (\ref{a10}) and (\ref{a12}) and obtain
\begin{equation}
\frac{\dot{\alpha}}{\alpha}=\frac{12}{\varepsilon}(j-q-2)H\frac{n(n-1)\lambda
x^{n-2}}{1-n\lambda x^{n-1}}
\end{equation}
\begin{equation}
\frac{\triangle \alpha}{\alpha}=n\lambda x^{n-1}(n\lambda
x^{n-1}+2)
\end{equation}
where $x\equiv R/R_{c}=\frac{6}{\varepsilon}(1-q)h^2$ with
$h=H(z)/H_{0}$. Using the observational fact that $H_{0}=72\pm8
~km^{-1}s~Mpc^{-1}=7.4\times 10^{-10}~yr^{-1}$ \cite{freedman},
then the experimental bounds (\ref{a2}) and (\ref{a3}) imply
\begin{equation}
\frac{\dot{\alpha}}{\alpha}\mid_{z=0}=12(j_{0}-q_{0}-2)\frac{n(n-1)\lambda
x_{0}^{n-2}}{1+n\lambda x_{0}^{n-1}}<~10^{-5}
\label{a122}\end{equation}
\begin{equation}
\frac{\triangle \alpha}{\alpha}\mid_{z=0.45}=n\lambda
x^{n-1}(n\lambda x^{n-1}+2)<~10^{-7} \label{a13}\end{equation}
where we have set $\varepsilon=1$ and the subscript $0$ represents
the present value of a quantity.  It is now evident that the
relations (\ref{a122}) and (\ref{a13}) act as algebraic
constraints on the parameters $\lambda$ and $n$ for given values
of the quantities $j_{0}$, $q_{0}$, $q(z=0.45)$ and $H(z=0.45)$.\\
To estimate these quantities, we use the gold data set for $j_{0}$
and $q_{0}$ which gives $j_{0}=2.75^{+1.22}_{-1.10}$ and
$q_{0}=-0.86\pm 0.21$ \cite{blan}.  On the other hand, the Hubble
parameter $H$ and the deceleration parameter $q$ are related by
\begin{equation}
H(z)=H_{0}~exp~[\int_{0}^{z} (1+q(u))d\ln(1+u)]
\label{h}\end{equation} If a function $q(z)$ is given, then we can
find the Hubble parameter in the redshift $z$.  Here we use a
two-parametric reconstruction function characterizing $q(z)$
\cite{wang}\cite{q},
\begin{equation}
q(z)=\frac{1}{2}+\frac{q_{1}z+q_{2}}{(1+z)^2}
\label{ee}\end{equation} where fitting the model to the gold data
set gives $q_{1}=1.47^{+1.89}_{-1.82}$ and $q_{2}=-1.46\pm 0.43$
\cite{q}. Using this in (\ref{h}) yields \cite{q}
\begin{equation}
H(z)=H_{0}(1+z)^{3/2}exp[\frac{q_{2}}{2}+\frac{q_{1}z^2-q_{2}}{2(z+1)^2}]
\label{e}\end{equation} At $z=0.45$, (\ref{ee}) and (\ref{e}) give
$q=0.12$ and $H=1.28 H_{0}$. As an illustration, the changes of
$\frac{\dot{\alpha}}{\alpha}$ and $\frac{\Delta \alpha}{\alpha}$
in terms of the parameters $\lambda$ and $n$ are plotted in Fig.1.
\begin{figure}[ht]
\begin{center}
\includegraphics[width=0.49\linewidth]{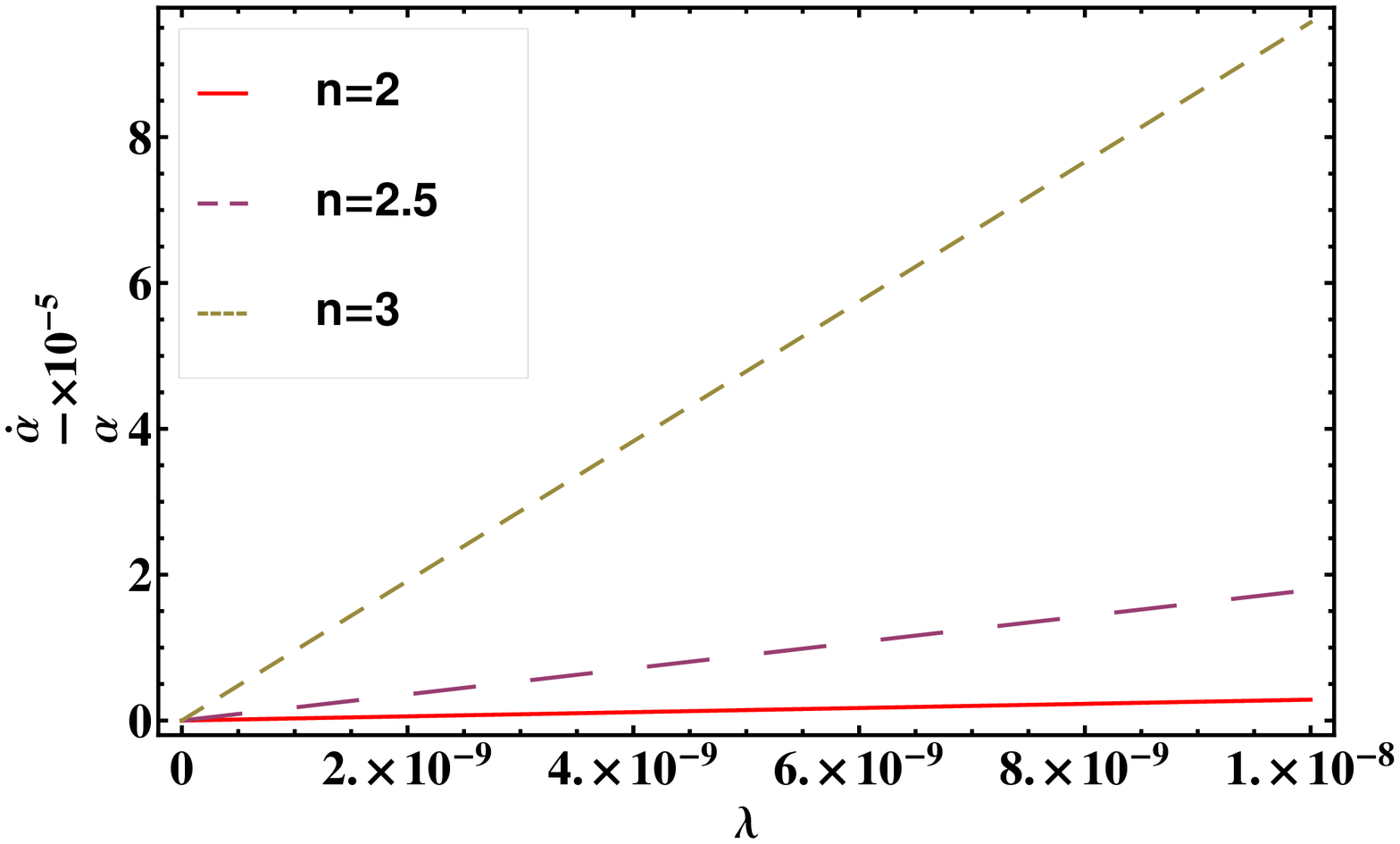}
\includegraphics[width=0.49\linewidth]{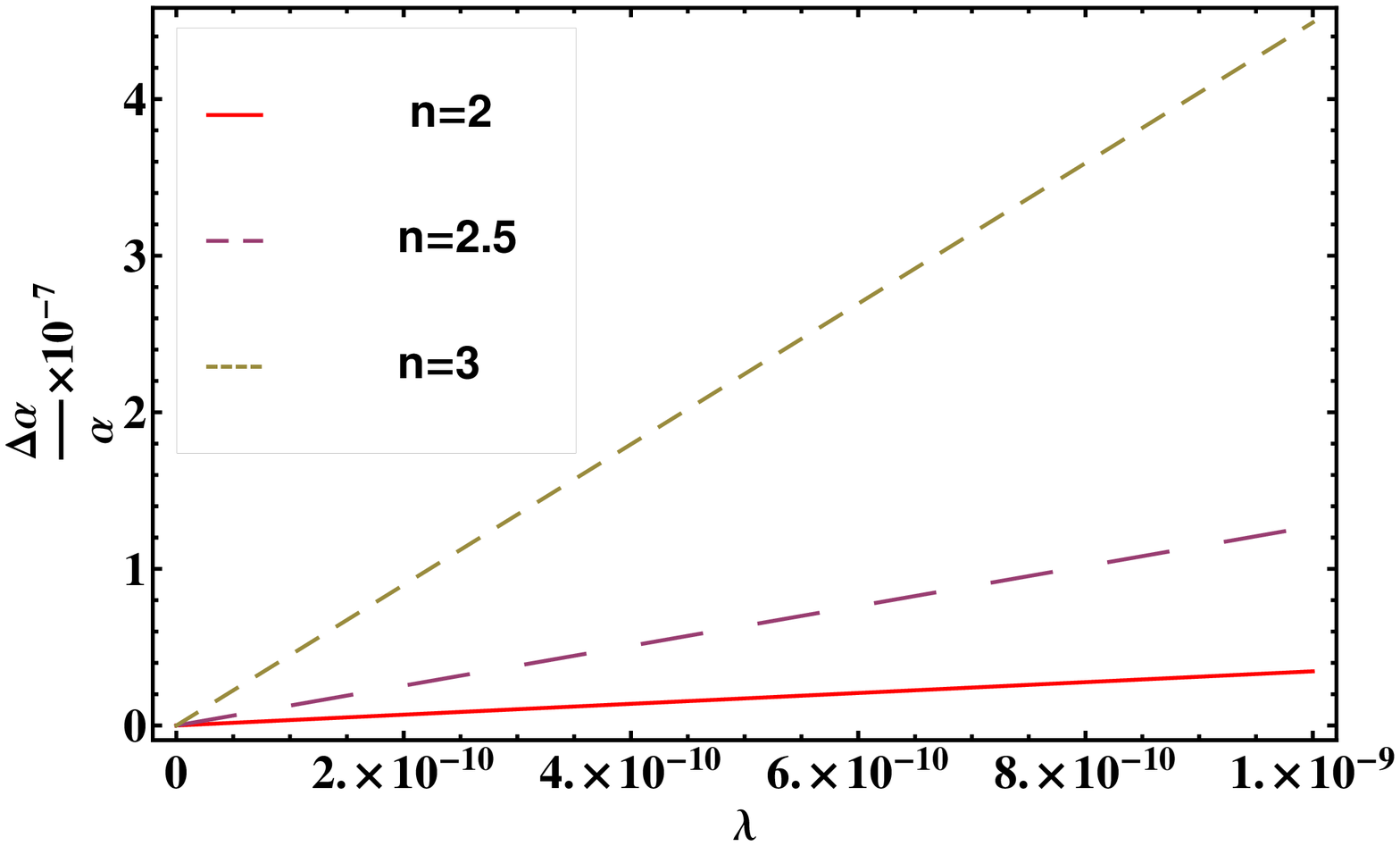}
\caption{The changes of $\frac{\dot{\alpha}}{\alpha}$ and
$\frac{\Delta \alpha}{\alpha}$ in terms of $n$ and $\lambda$ for
the model (\ref{R}). The vertical axes are scaled according to the
observational bound. The figures show that for larger values of
$n$ the slope of the lines increase.  This implies that for
moderate values of $n$, $\lambda$ should be extremely
small.\label{wzxy}}
\end{center}
\end{figure}
The figure shows that the bounds (\ref{a122}) and (\ref{a13}) are
satisfied for small values of the parameters.  For instance, for
$\lambda$ to be of the order of unity one should have $n<
10^{-4}$.  A moderate value of $n$ requires an extremely small
value for $\lambda$ which makes the model (\ref{R}) to be hardly
distinguishable from Einstein
gravity.  \\
Now we consider the models presented by Starobinsky \cite{star}
\begin{equation}
f(R)=R-\lambda R_{c} \{1-[1+(\frac{R}{R_{c}})^2]^{-n}\}
\label{s}\end{equation} and Hu-Sawicki \cite{hs}
\begin{equation}
f(R)=R-\lambda
R_{c}\{\frac{(\frac{R}{R_{c}})^n}{1+(\frac{R}{R_{c}})^n}\}
\label{hs}\end{equation} which can be written in a unified form
\cite{m}
\begin{equation}
f(R)=R-\lambda R_{c}
\{1-[1+(\frac{R}{R_{c}})^n]^{-\frac{1}{\beta}}\}
\label{hss}\end{equation} This parametrization corresponds to
(\ref{s}) and (\ref{hs}) with $n=2$, and $\beta=1$, respectively.
To apply the experimental bounds, we use (\ref{a10}) and
(\ref{a12}) and obtain
\begin{equation}
\frac{\dot{\alpha}}{\alpha}\mid_{z=0}=12(j_{0}-q_{0}-2)\frac{\frac{\lambda
n}{\beta}
x_{0}^{n-2}(1+x_{0}^n)^{-\frac{1}{\beta}-1}[n(\frac{1}{\beta}+1)x_{0}^n
(1+x_{0}^n)^{-1}-(n-1)]}{1-\frac{\lambda n}{\beta}
x_{0}^{n-1}(1+x_{0}^n)^{-\frac{1}{\beta}-1}}~<~10^{-5}
\label{a14}\end{equation}
\begin{equation}
\frac{\triangle \alpha}{\alpha}\mid_{z=0.45}=\frac{\lambda
n}{\beta}x^{n-1}(x^n+1)^{-\frac{1}{\beta}-1}\{\frac{\lambda
n}{\beta}x^{n-1}(x^n+1)^{-\frac{1}{\beta}-1}-2\}~<~10^{-7}
\label{a15}\end{equation} We have used the same data set for the
quantities $j_{0}$, $q_{0}$, $q(z=0.45)$ and $H(z=0.45)$ and
plotted (\ref{a14}) and (\ref{a15}) in Fig. 2.
\begin{figure}[ht]
\begin{center}
\includegraphics[width=0.49\linewidth]{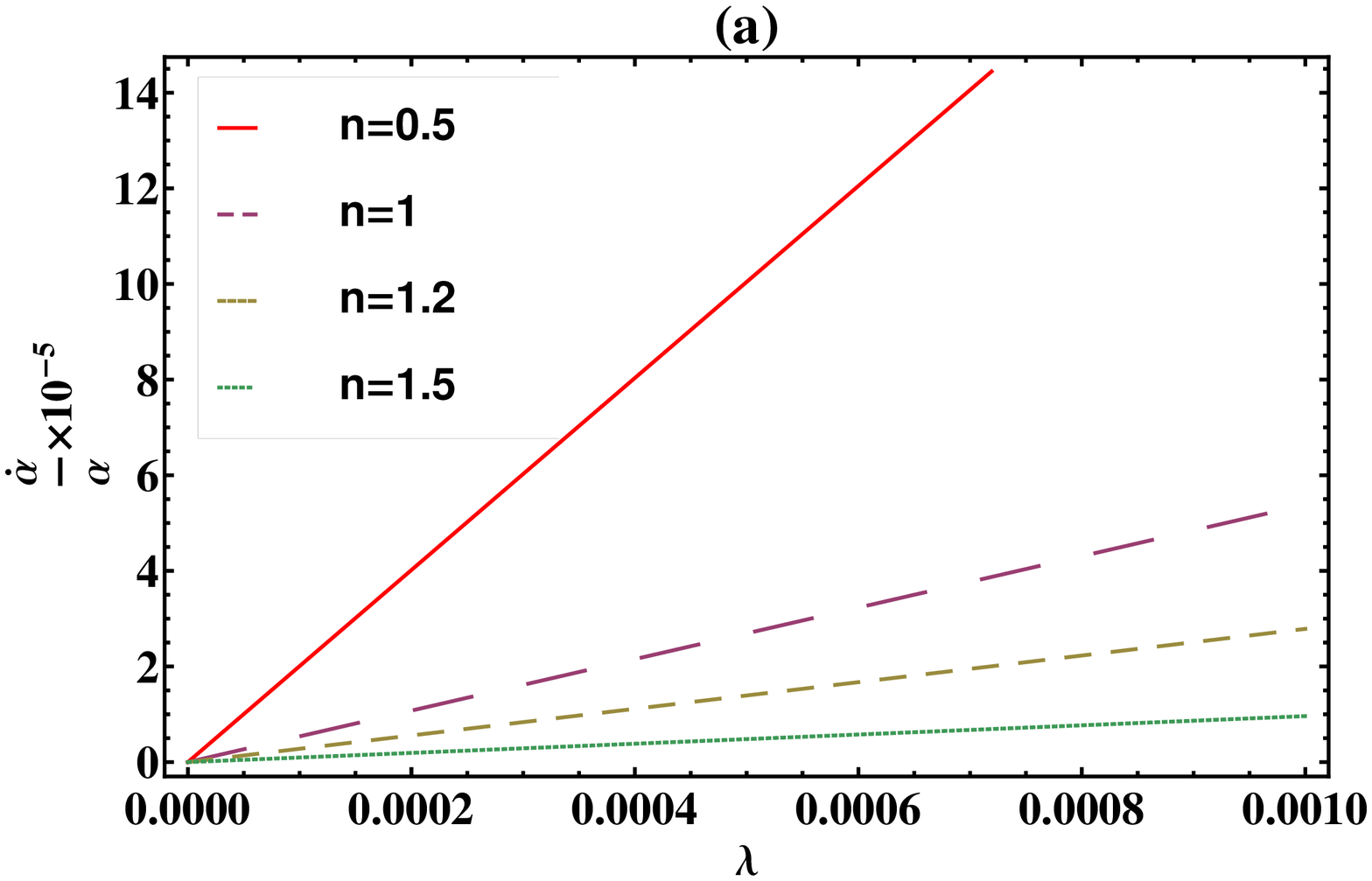}
\includegraphics[width=0.49\linewidth]{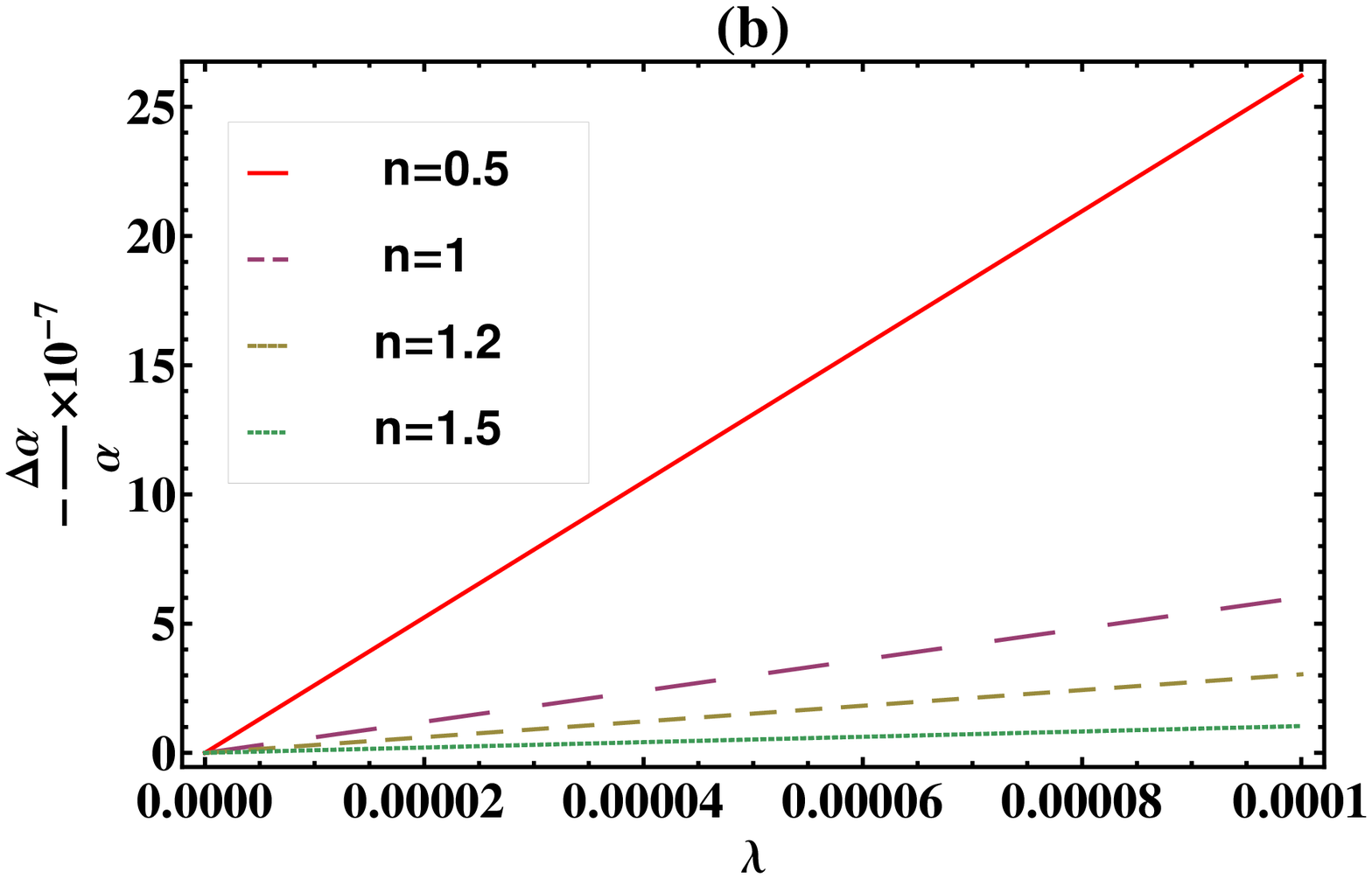}
\end{center}
\end{figure}
\begin{figure}[ht]
\begin{center}
\includegraphics[width=0.49\linewidth]{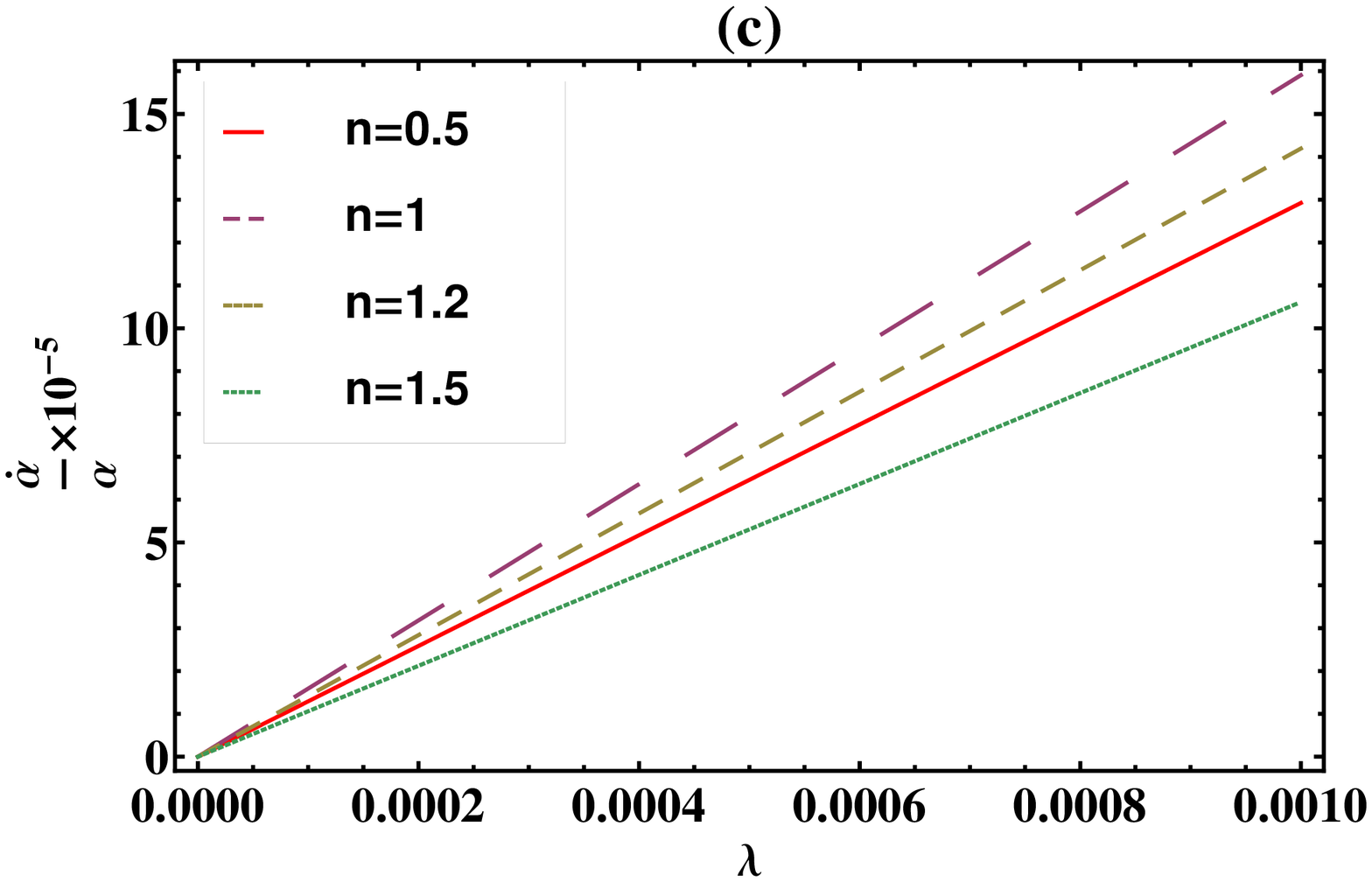}
\includegraphics[width=0.49\linewidth]{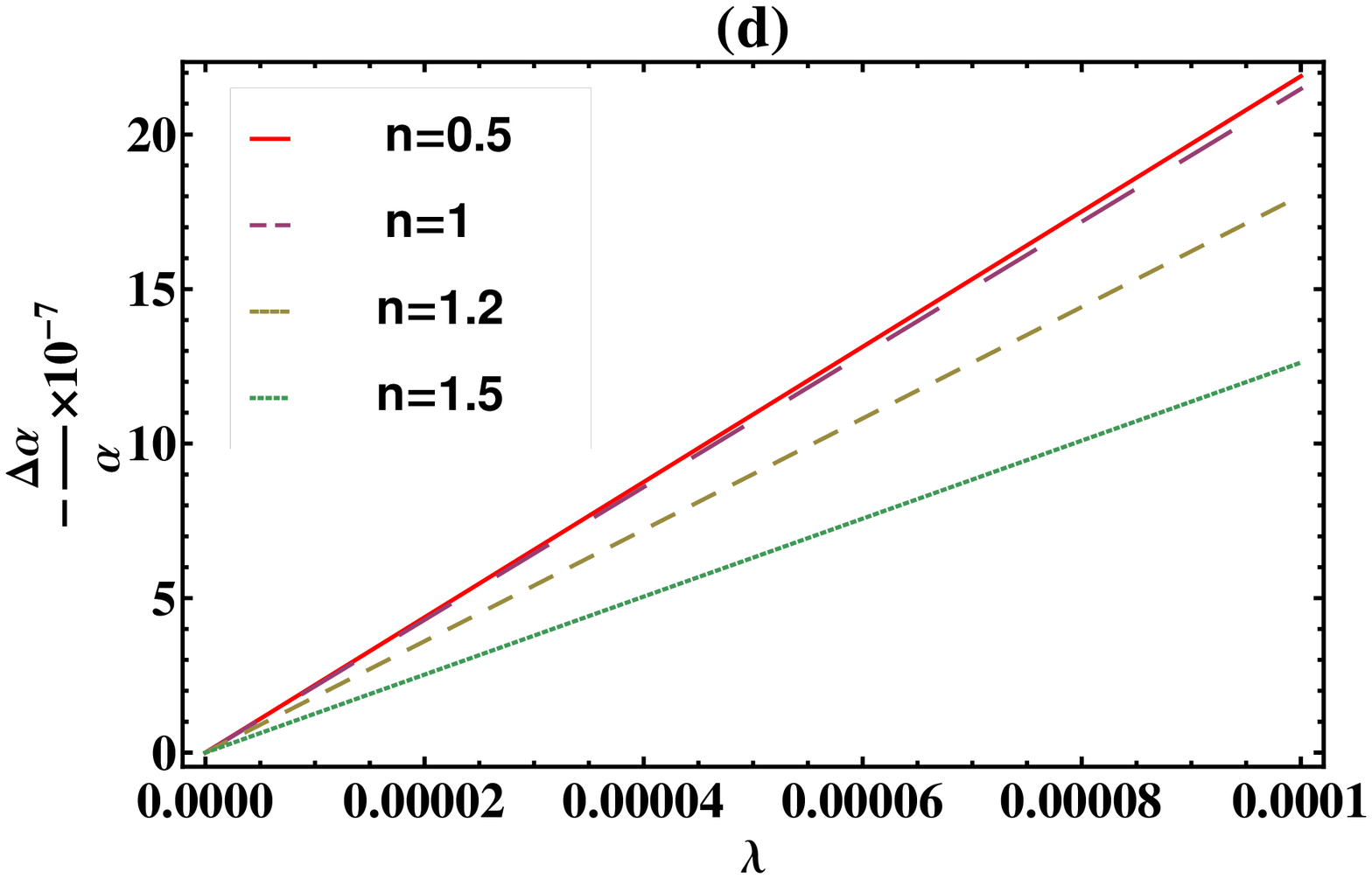}
\caption{The changes of $\frac{\dot{\alpha}}{\alpha}$ and
$\frac{\Delta \alpha}{\alpha}$ for the Starobinsky (panels a, b)
and Hu-Sawiki (panels c, d) models. The vertical axes are scaled
according to the observational bound.  In contrary to the model
(\ref{R}), in these models the larger values of $n$ correspond to
decrease of the slope of the lines. \label{wzxy}}
\end{center}
\end{figure}
The bounds satisfy in the region of the parameter space which
corresponds to larger values of $n$ and smaller values of
$\lambda$. For a given value of $n$, the acceptable range of
$\lambda$ is $[0, \lambda_{1}]$ with $\lambda_{1}$ being the
maximum value of $\lambda$ which satisfies the experimental
bounds. If we take larger values of $n$ the bounds (\ref{a14}) and
(\ref{a15}) allow a larger $\lambda_{1}$ which means that the
whole range of acceptable $\lambda$ is larger.  The figures show
that for a given $n (n>1)$, $\lambda_{1}$ for the model (\ref{hs})
is smaller than that for the model (\ref{s}).  For $n$ near unity,
the acceptable values of $\lambda$ are very small for the both
models.

~~~~~~~~~~~~~~~~~~~~~~~~~~~~~~~~~~~~~~~~~~~~~~~~~~~~~~~~~~~~~~~~~~~~
\section{Conclusion}
A possible modification of gravity is replacing the curvature
scalar in the Einstein-Hilbert action by a general $f(R)$
function.  It is well-known that these $f(R)$ gravity models are
conformally equivalent with a class of quintessence models in
which gravity is minimally coupled with a scalar field with some
appropriate potential function.  In this Einstein frame
representation of $f(R)$ models, conformal transformations induce
a coupling of the scalar field with
ordinary matter which may potentially lead to variation of constants of nature.\\
Although both quintessence models and $f(R)$ theories in the
Einstein frame can lead to changes of fundamental constants, there
is a basic different between these two alternatives.  In the
quintessence models, the coupling of the scalar field with matter
system is characterized by an arbitrary gauge function $A(\phi)$.
Cosmological evolution of $\phi$ is then determined by $A(\phi)$
and some potential function of the scalar field $V(\phi)$. On the
other hand, we have shown that in $f(R)$ theories the possible
variation of constants only depends on the functional form of
$f(R)$.  This is due to the fact that the gauge function is fixed
by conformal transformations and is given by an exponential
function and the potential of the scalar field is determined by
$f(R)$ itself.\\
Based on these arguments we have introduced a criterion that can
be used to probe the cosmological viability of $f(R)$ theories. In
this criterion, we have considered a general $f(R)$ theory in the
Einstein frame and restrict ourselves to temporal variations of
fine structure constant.  We have shown that for any $f(R)$
gravity model variations of fine structure constant can be written
in terms of some cosmological parameters (such as $q$, $j$ and
$H$) in an appropriate redshift.  In particular, the relations
(\ref{a10}) and (\ref{a12}) allows us to apply the experimental
bounds on the rate and the relative changes of $\alpha$ for any
$f(R)$ gravity model.  We have then used the gold data set to
estimate $q$, $j$ and $H$ at redshifts $z=0$ and $z=0.45$ for applying to some particular $f(R)$ functions.\\
Specifically, we have applied the above procedure to models of
Starobinsky and Hu-Sawicki.  We have shown that these
two-parameter models satisfy the experimental bounds in the
regions of parameter space which correspond to larger values of
$n$ and smaller values of $\lambda$.  This result is more
restrictive than the constraints coming from local gravity
experiments which set a lower bound only on $n$ \cite{c}.  In our
analysis both the parameters $n$ and $\lambda$ should satisfy the
algebraic constraints (\ref{a14}) and (\ref{a15}) so that a
particular value for one parameter affect the whole of range of
validity of the other. Moreover, we have shown that for a given
value of $n$, there is an upper bound on the allowed range of
$\lambda$ ($\lambda_{1}$). For $n=1$, $\lambda_{1}$ will be of the
order of $10^{-4}$ in both models.  It is important to note that
the local gravity  analysis gives $n>0.5$ as the allowed range of
$n$ \cite{c}.  Although the outcome of our analysis is consistent
with this result, it however emphasizes that this parameter should
take values sufficiently larger than unity in order that the
models keep an appropriate departure from Einstein gravity.


\end{document}